# Where Do All These Search Terms Come From? – Two Experiments in Domain-Specific Search

Daniel Hienert and Maria Lusky

GESIS – Leibniz Institute for the Social Sciences
`daniel.hienert@gesis.org, maria.lusky@gmail.com`

**Abstract.** Within a search session users often apply different search terms, as well as different variations and combinations of them. This way, they want to make sure that they find relevant information for different stages and aspects of their information task. Research questions which arise from this search approach are: Where do users get all the ideas, hints and suggestions for new search terms or their variations from? How many ideas come from the user? How many from outside the IR system? What is the role of the used search system? To investigate these questions we used data from two experiments: first, from a user study with eye tracking data; second, from a large-scale log analysis. We found that in both experiments a large part of the search terms has been explicitly seen or shown before on the interface of the search system.

## 1 Introduction

For simple information needs users can enter some keywords into the search bar and most of the times receive the right answer. However, for more complex information needs users tend to vary their search terms, add new terms or use combinations of them in order to achieve better results and to uncover new aspects of an advanced information problem. This scenario of searching information is a rather complex one as we have an interplay between the user, the search system and information outside the search system, e.g. in other online or offline sources. Input for new search attempts can therefore be derived from several sources and may additionally be subject to cognitive processes by the user.

A first set of research questions therefore is: What are the sources of new search terms? What is the share of input coming from the user, the search system or other sources? Where and when in the search process are potential new search terms recognized? Further research questions are: How long does it take until a potential term is used in a search? And which cognitive processes are applied on it? The answers to these questions have implications for the design of our search systems. They tell us where, when and how in the search process users are getting ideas for new search terms. This can be a basis for designing new supporting services within a search system that help users in the right place at the right time of the search session.

To answer the basic question where and when users get ideas for new search terms from, we use data from two related experiments in the field of social science literature search: (1) a task-based user study with 32 subjects and recorded eye tracking data, and (2) a large scale log analysis with log data of nine years. The first experiment will tell us explicitly if users have seen new search terms in their search process before they use them. The second experiment can tell us on a large scale if new search terms have been shown on the system before being used by the user.

## 2   Related Work

In this section we will present related work on interactive search models, evaluation models and the analysis of search terms used in a search session.

### 2.1   Models for information search

The classical Cranfield paradigm is a rather technical model with the goal to optimize search results for a given query. Interactive Information Retrieval (IIR), in contrast, tries to incorporate the user into the search process and explicitly take into account the interactivity between the user, the system and the content. The IIR evaluation model of Cole et al. [2] for example models the search process by starting at a problematic situation a user is facing, which triggers the overall goal and the task to seek information with different seeking strategies to solve the issue. Another framework for IIR is the IPRP model [4] which sketches the search process as transitions between situations, where the user can choose in each situation from a list of choices. Another search model is exploratory search [12] which explicitly addresses the case of a user who is not only looking up a simple information fact, but who is engaging in a more complex problem or unknown area and who is learning and investigating, trying to understand the problem a bit better step by step in his search process.

### 2.2   Evaluation methods

For the evaluation of IIR systems and situations, different methods can be used. IR evaluation for a long time has focused primarily on the system view. However, user studies can give valuable insights on how users interact with IR systems. Kelly [11] gives a good overview of user-oriented evaluation methods for IIR. Advantages of these kinds of studies are that real users are observed (maybe within a given task) and the way they interact with the system. These methods enable us to investigate the information seeking behavior of users on the one hand and how an IR system can support users (or hinder them) to gain new insights on the other hand. Disadvantages of user studies are that they are often costly, small-scaled and their significance can therefore be limited.

Eye tracking as a method in IIR evaluation can be used for various purposes. First, it shows the user's attention to different parts of the IIR system's interface, e.g. the search bar or an item on the result page. For example, the F pattern is known as a regularity of how users read web pages [13]. Second, it shows which kinds of texts (title, abstract etc.) users are scanning and how they do it. Longer dwell times can e.g. indicate the user's interest in an item. Third, eye movement patterns can reveal cognitive representation of information acquisition and were used to derive user groups of different domain knowledge and working on different search tasks [3]. The E-Z Reader model [15] assumes that text reading is a serial process with the user's attention to one word after the other. Each of these attention spots is called a fixation. A jump from one fixation to the next one is called a saccade. Within a fixation the E-Z Reader model divides the process of understanding the word meaning (lexical processing) in two stages L1 and L2. The first stage L1 describes the "familiarity check" – the basic word identification – which can be processed with a maximum mean time of 104ms [16]. With the end of this stage the programming of the saccade to the next word is initiated. The second stage L2 ends with the full understanding of the word. Both stages take an overall time from 151ms to 233ms on average [15]. The time for lexical processing depends on a number of variables such as the word length, the word frequency in a language corpus and the word/text difficulty [15].

Log analysis as an evaluation methodology in IIR stands in the middle between user- and system-oriented studies. Log analysis can capture user interaction with the system on a large scale, however, it cannot anticipate the user's information need, the task, the overall problem, the situation and context of the search [9]. It is important to distinguish between web search engine log analysis and digital library (DL) log analysis [1]: in web search retrieved documents are web pages; in DL search documents are maintained by information professionals and are often organized by knowledge organization systems. Also, DL search is often specific for a certain domain, community or topic.

### 2.3 Analysis of search term usage

The focus in IIR on interactivity also suggests having a deeper look at the whole search process. Thereby the event(s) of a user entering keywords into the search bar is certainly important. Transaction log analysis (TLA) has already dealt with different *statistical measures* of search term usage for a long time [14]: How many search terms were used? How long are search terms on average? In this sense a lot of studies were conducted in different domains (e.g. for Pubmed users [7]). Along that, users of *different domains* search differently: for example for the domains of history and psychology see [19]. On the one hand the *effectiveness* of different sources of search terms had been investigated, especially the use of a controlled vocabulary from a thesaurus vs. free uncontrolled terms [17]. Another aspect are the *patterns of query reformulation*: In which way do users add, delete and replace query terms? For example, Jansen et al. [8] found that generalization and specialization are main transition patterns in web search. Jiang & Ni [10] recently studied what affects word changes in query reformulation based on word-, query- and task-level.

So far, in research only little attention has been given to the *sources of search terms*. Spink & Saracevic [18] conducted a "real-life" study with academic users from several domains and identified five sources of search terms: (1) the question statement the subjects had to fill out with their own information problem, (2) user interaction, (3) a thesaurus, (4) an intermediary and (5) the retrieved items. Yue et al. [20] did a smaller work investigating where query terms come from in collaborative web search. We build up on this research and investigate if users have explicitly seen search terms before applying them in a free search. In a large-scale experiment we check if search terms have been shown on the system before being used.

## 3  Evaluation Context

In this section we first briefly describe the evaluation system, a real-world digital library for social science literature information. Then we report on the typical search processes in the search system to understand what users' possibilities are for getting search term suggestions.

### 3.1  System Description

Sowiport [5] is a digital library for social science information with more than nine million bibliographic records, full texts and research projects. The portal gives an integrated search access to twenty German and English-language databases. About 25,000 unique visitors per week are visiting the portal, mainly from German-speaking countries. One of the services for supporting users in their search process is the Combined Term Suggestion Service (CTS) [5]. When the user enters characters into the search bar, the service proposes different term suggestions: (1) auto completion terms from the thesaurus for the social sciences, (2) related, broader and narrower terms from the thesaurus, (3) statistically related terms from a co-occurrence analysis based on titles and abstracts, and (4) author names based on auto completion.

### 3.2  Search Process

The search process in Sowiport normally follows regular patterns which already were visualized and analyzed with the WHOSE toolkit [6] and which are comparable to the ones in other literature information systems. A first possibility is that users enter Sowiport via the homepage. They can then directly initiate a search via the search bar, where term recommendations from the CTS are shown. The user can also switch to the advanced search form and start there. The next step is the result page which shows a list of twenty documents with title, authors, source and a highlighting text fragment that shows the textual context where the user terms were found in the document. Each document has (where available) links to Google Scholar, Google Books and to the full text (via DOI or URL directly to the journal, proceedings, archive, university or personal websites). Users can follow these links and read (parts of) the full text outside the Sowiport system. On the result page, users can continue and refine their search by paging, choosing from the facets, entering new search terms, or starting a new search for persons, proceedings or journals from the metadata of each record. If one of the records seems to be relevant, the user can enter the detailed view with a click on the title. Then, all metadata entries such as title, source, categories, topics, abstract, references and citations are shown. From here, the users can continue by choosing from similar or related records on the left page section, by choosing a document from references or citations, by entering new search terms in the search bar above or by initiating a new search by clicking on the metadata entries. A large part of users enters Sowiport through a detailed view of a record coming directly from a search engine. These users can then continue their search process with the options of the detailed view.

We can distinguish between two possibilities of how users can initiate a new search process: (1) by simply clicking on a link. This can be done in the result list for authors, proceedings, journals and from the facet section and in the detailed view for all metadata of the record (authors, keywords, categories, journal, proceeding) or (2) by manually entering new search terms into the search bar. This can be done in the search bar on the home page, in the advanced search form and always in the search bar above the result list and the detailed view. *In this paper we will focus on where users get ideas and suggestions for new search terms from when entering them freely in a search form* (for brevity we call it in the following a "*free search*") as here users explicitly enter new search terms which come from the user's mind (and are not readily prepared by the system).

Suggestions for new search terms can come on the *system side*: (1) from the search term recommender when entering terms in one of the search forms, (2) on the result page from titles, authors, sources and highlighted fragments of each search result, (3) from the facet section shown on the result page on the left, (4) from the detailed view which shows all fields such as title, source, categories, topics, abstract, references and citations. Additionally, search terms can derive from (5) the full text which is checked typically *outside the retrieval system* and finally (6) from the *user side* who may have some keywords on his mind, a list of references printed out on his desk or printed text with markers here and there.

## 4    Experiment I: User Study

For a first investigation we used data from a user study. For each free search we investigated if the search term was seen by the user on the search system by using eye-tracking data.

### 4.1    Description

We used data from a lab study with two groups of 16 subjects each (20 female, 12 male) that took place in single sessions with a duration of 30 minutes. While one group consisted of bachelor and master students, the other group comprised only postdoctoral researchers. All subjects worked in different fields of the social sciences. The students were between 22 and 35 years old (m=26.38, sd=3.76), while the age of the postdocs ranged between 30 and 62 (m=40.19, sd=9.23). On a 5-point Likert scale (1="very rarely", 5="very often"), the subjects rated their frequency of use of digital libraries on average with 2.78 (sd=1.02) and of Sowiport with 2.22 (sd=1.14). They also considered their search experience in digital libraries as moderate (m=2.91, sd=0.91).

All subjects were given the same document about the topic "education inequality", opened in Sowiport, and were asked to find similar documents using our digital library. To do so, they had a total time of 10 minutes. During the task their eye movements as well as the screen were recorded. We made sure the conditions were the same in each session: The subjects used a mouse, a keyboard and a 22″-monitor connected to a laptop. The laptop display served as an observation screen. All subjects worked with Mozilla Firefox. For tracking their eye movements we used the remote eye tracking device *SMI iView RED 250* that was attached to the bottom side of the stimulus monitor. We calibrated the eye tracker with each subject using a 9-point calibration with a sampling frequency of 250Hz and only then started the experiment. For creating the eye tracking experiment as well as analyzing the gaze data, we used the corresponding software *SMI Experiment Suite 360°*.

### 4.2 Methodology

For analyzing the subjects' eye movements we created a gaze replay video for each subject, showing their scan paths during the whole session in order to determine the individual words the subjects looked at. The eye tracking software enabled us to make full screen records that also captured the navigation bar of the web browser and dynamic elements like the search term recommender. We used a fixation time threshold of 104ms as the beginning of the L2 period when the user starts to semantically understand the word. Since the user study was limited to the interaction between the user and our search system, these are the only two sources where search terms could be derived from. Therefore, we first detected each time a subject conducted a free search during the experiment and captured the search terms that were used. In a second step, we carefully observed the subject's scan paths of the session and checked if they had read the search terms before.

### 4.3 Results

The analysis of the gaze replay videos shows that for this task users are scanning through the result lists and detailed views looking for information that can help to solve the task. As a starting point they especially scan the metadata of the seed document, its references, citations and related entries. They use the title, keywords, abstract, references and citations to browse to related documents and conduct new searches. Terms for free searches were seen explicitly on the result list, in the detailed view or in other parts of the system. Table 1 shows the detailed results. The users conducted 82 free searches. About *78%* of user search terms were seen explicitly on the system before being used for a free search. The largest part comes from the detailed view (51.22%), then from the CTS (9.76%), the result list (4.88%), the references (4.88%), from related entries (4.88%) and from the thesaurus (2.44%). Metadata fields from which search terms were taken are the title (58.93%), keywords

(28.57%), abstract (7.14%), authors (3.57%), and categories (1.79%). In 21.95% of the cases the used search term had not been seen by the user prior to the search, which means that the search term was formed by the user. The diagrams in Figure 1 also show that the student and the postdoc group have very similar results.

In a lot of cases the terms later used for a free search query were seen by the user several times during the session. We measured an average time of 3:44 minutes from first sight to search and an average time of 1:27 minutes from last sight to search.

One third (29.27%) of the participants conducted cognitive operations of the terms seen. We identified the following categories: (1) *translation* (e.g. from German to English), (2) *separation* of compound terms and then taking only one part of the term for searching, (3) *nominalization* of terms from e.g. personifications to substantives, (4) *merging* of two terms seen and (5) *broadening* of terms.

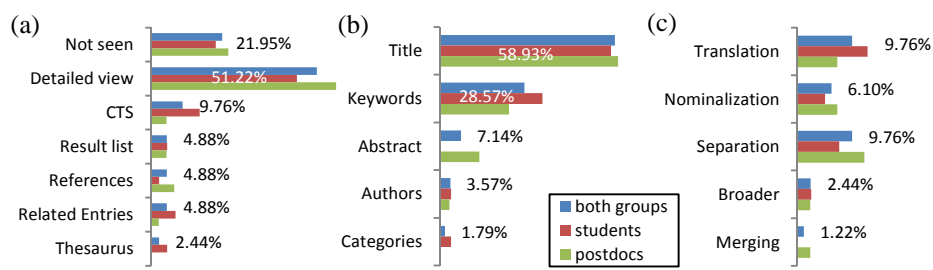

**Fig 1.** User study: (a) sources and (b) metadata fields where the search terms were seen and (c) the distribution of cognitive operations.

## 5 Experiment II: Log Analysis

In this second experiment we used the insight from the first experiment and wanted to find out on a large scale if applied search terms in a free search were shown before on the system. We used a log-based approach and computed for every free search if the used search term had been shown before in the session. Here, the investigation of search term sources was limited to the system side.

### 5.1 Dataset

For this experiment we used nine years of Sowiport's log data from between November 2007 and July 2016. The data derives from two different technical systems underlying Sowiport and from different sources, such as log files and logs in database tables. The dataset was cleaned from bots and search engines.

We extracted two user actions from the log data to a user action table: (1) A search action ("*search*") with the database fields session-id, timestamp, search form type (simple, advanced, URL), search field type (all, author, keyword, title, location, date, institution journal/proceedings, topic-feed), the user search terms and result list ids. A *free search* based on keywords (not persons, numbers, locations etc.) can then be identified from the action table by having the search form type set to "simple" or "advanced" and the search field type set to "all" or "keyword" and the user search terms not being empty. (2) A view record action ("*view_record*") with the fields session-id, timestamp and the doc-id of the viewed record.

This dataset was further filtered on the session side to (1) user sessions which either had at least one document view before a free search or (2) to sessions with at least two free searches with distinct user terms. In this kind of sessions the user had the chance to recognize a search term from the document view before or to learn from the system's output between two searches. The final evaluation dataset includes 96,067 user sessions with 602,065 searches and 523,638 record views. A single session contains on average 12 user actions and is about 16 minutes long.

### 5.2 Methodology

We built an algorithm that takes each individual user session and goes through each action, step by step in temporal order. For each session step we collected the metadata of the records which had been shown on the system in a collector. The metadata was cleaned from German and English stop words and stemmed to facilitate the comparison to user search terms later on. For a search action we collected the metadata of the result list entries (title, persons, keywords, categories). For a view_record action we collected the metadata of the viewed record (title, persons, keywords, categories, abstract). References and citations for that record would only be added to the collector if the user had clicked the appropriate tab in the user interface. Some information shown on the system were not collected, because it would have been too costly to compute them for each single search and record view. This affects namely the facet section on the result list and the highlighting fragments for each record that show in which context the user's search terms were found. For the detailed view we left the similar and related documents out of computation.

For each search action, the algorithm first checked if the search terms were taken from the term recommender. If not, it checked if the (stemmed and stop-word cleaned) search terms were shown in a previous session step by comparing them to the collected metadata. Therefore, it went backwards through the session, starting from the search event. Then each search term was compared to the metadata fields in the collector. The ordering of different metadata fields (title, keywords etc.) in the collector had an influence on the field in which the user term is found, because the user term was first checked against the first entry, then the second and so on. We chose the order of the user study (see Figure 1) as an empirical basis. For each hit, the session step, the source, document and metadata field where the term was found and the search term itself were recorded.

### 5.3 Results

Figure 2 shows the results of the log experiment. A share of *38.29%* (215,376 of 562,426) user search terms were shown by the system before being used in a free search. The source was in most cases (25.02%) the result list, then the detailed view

(13.27%), followed by the term recommender CTS (2.9%) and marginally the references (19 times - ~0%). Metadata fields, where search terms were derived from are keywords (57.13%), title (18.45%), persons (10.38%), abstract (8.45%) and categories (5.58%). We also measured the distance between the search action and the step in which the search term was shown on the system. Figure 2(c) shows that a large part (29.59% of 38.29% maximum) was shown within three steps, which is quite near the search action. Within 10 steps almost all search terms that were used were shown on the system (35.79% of 38.29% maximum). There are on average 2:30 minutes/9.35 session steps between first occurrence and the search and 2:04 minutes/3.64 session steps between last occurrence and the search step. On average, a term was shown 8.76 times within a session before being used in a free search.

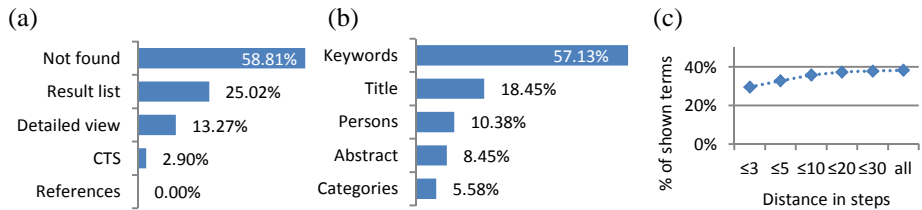

**Fig 2.** Log analysis: (a) sources and (b) metadata fields where the search terms were shown on the system and (c) the distance to the search action in session steps.

## 6   Discussion

The two different experimental approaches in our case have well completed each other. The user experiment visualized the process that users are explicitly scanning the user interface for information and in particular showed that in their free searches users apply terms they have seen before on the search system. Here, two different sources – system and user – were examined as possible sources of search terms. The log experiment then concentrated on the system side as a source for search terms and checked if there is a regularity.

In the user study a large part (78%) came from the system and was seen; the rest came from the user and other sources. This really high value can be surely ascribed to the specific evaluation task. We additionally experimented with lower and higher fixation times. With a fixation time of 50ms some more search terms had been recognized before the search, with 151ms some less, but the core of search terms which were seen was stable.

In the log analysis we found a value of about 38% of terms that were shown before being used in a free search. This is still a high value, but surely based on a different kind of user population with a diversity of tasks and topics. In the log analysis we can only assume that the users have explicitly seen the terms. However, the identified scan process in the user study, the number of search terms occurrence in the session prior search and the scale of the experiment in the log analysis indicate a high probability for this being true.

In both experiments a considerable amount of free search terms originated from different parts of the system, which should give system designers a higher responsibility to support users in finding the right terms. Support has to be given not only via a typical term recommender (which has been long-time acknowledged in our field), but also in all steps of the search process, as well as while viewing the entries in the result list and checking a record in detail.

In terms of system and user sources, Spink & Saracevic [18] in their experiment found that user interaction was responsible for 23% of the search terms, while 11% came from Term Relevance Feedback [the rest came from the question statement (38%), thesaurus (19%) and intermediary (9%)]. Certainly, our and their results are hard to compare, because of the different settings of the experiment. However, on the system side they have focused on a relevance feedback loop, in which users chose terms from documents they found relevant. This is in contrast to our experiment, where we take into account the *whole* search system as a source for new search terms.

In detail, in both experiments suggestions for search terms had been taken from the detailed view (51.22% and 13.27%), the result list (4.88% and 25.02%), from the term recommender (9.76% and 2.90%) and other sources. This again shows that interesting new keywords are extracted at different steps of the search process. A typical term recommender is only one of several sources where users are taking ideas from for new free searches. Metadata fields where search terms were taken from were relatively similar in both experiments. Most came from the keyword section (28.07% and 57.13%) and the title (59.65% and 18.45%), from the abstract (7.14% and 8.45%), persons (3.57% and 10.38%), and categories (1.79% and 5.58%).

Following the search processes in the user experiment showed that search terms were shown several times in the system before users applied them in a free search. In the log analysis, applied search terms had been shown in the system up to eight times before being used. Although both experiments had different kinds of tasks (exploratory search in the user experiment; a diversity of tasks in the log analysis), the time spans from first sight and last sight until search are comparable. It took about 3:44/2:30 minutes from first sight and 1:27/2:04 minutes from last sight to the search event. Additionally, the log experiment shows us that the largest share of terms were shown within three session steps – thus from an interaction perspective really near the search action.
All in all, by taking into account the whole search system, we can see that steps in the session beforehand influence the actual step, which is a strong argument for the whole session or interactive information retrieval discussion.

## 7     Conclusion & Future Work

In this paper we conducted two experiments to investigate where users are taking ideas and suggestions for new search terms in free searches from. The user experiment showed well the process of scanning information and taking term suggestions from the system that have been shown at different sources, such as the result list, the detailed view or the term recommender. The log analysis showed on a large scale that one third of search terms had been shown on the system before the users conducted a search query with these terms. Answering our research question from the beginning, we can say that a good share of search terms comes from the system. The other parts are information from outside the system, but from online sources (e.g. reading full texts or articles in another tab) and from the user side with printed texts, ideas from discussions etc.

Search terms were seen and shown up to eight times in the search session and it could take some minutes until they were used in a free search. This again shows that the segmentation of the search process to query-response is too short-sighted, but user perception in the process minutes before querying can massively influence the actual action step. This also somehow negates user models with the assumption that the actual step is only influenced by the action before. The user experiment also showed that users are conducting cognitive processing of seen terms such as translation or separation.

We can conclude that finding new search terms is a process: (I) A good share of new free search terms comes from the system. (II) Search terms are shown and seen several times on the system before being used. (III) Terms can come from different parts of the system and from different metadata fields. (IV) Search terms are seen at different points in time within the session and it can take some time until they are used. (V) New search terms partly underlie cognitive operations from the user.

This research shows that searching and especially finding new free search terms is a complex process with interaction between the user, the system, the content and other entities online and offline. The user's state is influenced by all parts of the system and the user influences the system's state. In future work we want to concentrate even more on examining which interaction processes happen within a whole search session and how we can develop more suitable user models that capture these processes.

**Acknowledgements:** This work was partly funded by the DFG, grant no. MA 3964/5-1; the AMUR project at GESIS. The authors thank the focus group IIR at GESIS for fruitful discussions and suggestions.

## References

1. Agosti, M. et al.: Web log analysis: a review of a decade of studies about information acquisition, inspection and interpretation of user interaction. Data Min. Knowl. Discov. 24, 3, 663–696 (2011).
2. Cole, M. et al.: Usefulness as the criterion for evaluation of interactive information retrieval. In: Proceedings of the Workshop on Human-Computer Interaction and Information Retrieval. pp. 1–4 (2009).
3. Cole, M.J. et al.: User Activity Patterns During Information Search. ACM Trans Inf Syst. 33, 1, 1:1–1:39 (2015).

4. Fuhr, N.: A Probability Ranking Principle for Interactive Information Retrieval. Inf Retr. 11, 3, 251–265 (2008).
5. Hienert, D. et al.: Digital Library Research in Action: Supporting Information Retrieval in Sowiport. -Lib Mag. 21, 3/4, (2015).
6. Hienert, D. et al.: WHOSE – A Tool for Whole-Session Analysis in IIR. In: Proceeding of ECIR 2015. pp. 172–184 Springer (2015).
7. Islamaj Dogan, R. et al.: Understanding PubMed user search behavior through log analysis. Database J. Biol. Databases Curation. 2009, bap018 (2009).
8. Jansen, B.J. et al.: Patterns and transitions of query reformulation during web searching. Int. J. Web Inf. Syst. 3, 4, 328–340 (2007).
9. Jansen, B.J.: Search log analysis: What it is, what's been done, how to do it. Libr. Inf. Sci. Res. 28, 3, 407–432 (2006).
10. Jiang, J., Ni, C.: What Affects Word Changes in Query Reformulation During a Task-based Search Session? In: Proceedings of the 2016 ACM Conference on Human Information Interaction and Retrieval, CHIIR 2016, Carrboro, North Carolina, USA, March 13-17, 2016. pp. 111–120 (2016).
11. Kelly, D.: Methods for Evaluating Interactive Information Retrieval Systems with Users. Found Trends Inf Retr. 3, 1—2, 1–224 (2009).
12. Marchionini, G.: Exploratory search: from finding to understanding. Commun. ACM. 49, 4, 41–46 (2006).
13. Nielsen, J.: F-Shaped Pattern For Reading Web Content, https://www.nngroup.com/articles/f-shaped-pattern-reading-web-content/.
14. Peters, T.A.: The history and development of transaction log analysis. Libr. Hi Tech. 11, 2, 41–66 (1993).
15. Reichle, E.D. et al.: E-Z Reader: A Cognitive-control, Serial-attention Model of Eye-movement Behavior During Reading. Cogn Syst Res. 7, 1, 4–22 (2006).
16. Reichle, E.D. et al.: Using E-Z Reader to simulate eye movements in nonreading tasks: a unified framework for understanding the eye-mind link. Psychol. Rev. 119, 1, 155–185 (2012).
17. Rowley, J.: The Controlled Versus Natural Indexing Languages Debate Revisited: A Perspective on Information Retrieval Practice and Research. J Inf Sci. 20, 2, 108–119 (1994).
18. Spink, A., Saracevic, T.: Interaction in information retrieval: selection and effectiveness of search terms. JASIS. 48, 8, 741–761 (1997).
19. Yi, K. et al.: User search behavior of domain-specific information retrieval systems: An analysis of the query logs from PsycINFO and ABC-Clio's Historical Abstracts/America: History and Life. J. Am. Soc. Inf. Sci. Technol. 57, 9, 1208–1220 (2006).
20. Yue, Z. et al.: Where Do the Query Terms Come from?: An Analysis of Query Reformulation in Collaborative Web Search. In: Proceedings of the 21st ACM International Conference on Information and Knowledge Management. pp. 2595–2598 ACM, New York, NY, USA (2012).